\documentclass[prl,twocolumn,superscriptaddress,showpacs]{revtex4}
\usepackage{graphicx}
\usepackage{epsfig}
\usepackage{amssymb,amsmath,amsfonts,hyperref}
\usepackage{wasysym}
\usepackage{latexsym}

\newcommand{\be}{\begin{eqnarray}}
\newcommand{\ee}{\end{eqnarray}}

\begin{document}
\title{Impurity spin texture at a deconfined quantum critical point}
\author{Argha Banerjee}
\affiliation{\small{Tata Institute of Fundamental Research, 1, Homi Bhabha Road,
Mumbai 400005, India}}
\author{Kedar Damle}
\affiliation{\small{Tata Institute of Fundamental Research, 1, Homi Bhabha Road, Mumbai
400005, India}}
\author{Fabien Alet}
\affiliation{Laboratoire de Physique Th\'eorique, Universit\'e de Toulouse, UPS, (IRSAMC), F-31062 Toulouse, France}
\date{February 6, 2010}

\begin{abstract}
The spin texture surrounding a non-magnetic
impurity in a quantum antiferromagnet is a sensitive probe of the novel physics of a class of
quantum phase transitions between a N\'eel ordered phase and a valence bond solid phase in
square lattice $S=1/2$ antiferromagnets. Using a newly developed $T=0$ Quantum
Monte Carlo technique, we compute this spin texture at these transitions and find that it does {\em not} obey
the universal scaling form expected at a scale invariant quantum critical point. We also identify the precise logarithmic form of these scaling
violations. Our results are expected to yield important clues regarding the
probable theory of these unconventional transitions~\cite{Senthil_etal_Science2004}. 
\end{abstract}

\vskip2pc

\maketitle

A particularly elegant strategy in the study of strongly
correlated materials exploits the presence of small concentrations of well-characterized impurities in an otherwise pure sample. Each impurity acts
more or less independently of the others to alter the state of the system around it, and these impurity-induced charge and
spin textures can then be picked up by nuclear magnetic resonance
(NMR) or scanning tunneling microscopy experiments. As these local responses are
characteristic signatures of the underlying low temperature state, such
experiments provide a valuable window to the underlying
physics, especially if the state in question has strong correlations but
no obvious charge or spin order.\cite{Alloul_etal_RMP2009} 

Some of these experiments~\cite{SpinExp} have focused on the effects of non-magnetic
impurity atoms which give rise to a missing-spin defect in strongly correlated Mott insulators. Due to the uncompensated Berry phase
associated with such a missing moment~\cite{Sachdev_etal_Science1999}, it induces a non-trivial pattern
of spin density around it, and a direct signature of this spin texture
can be obtained by analyzing the pattern of Knight shifts in
NMR experiments.
Other experiments have also studied such effects in cuprate high-T$_c$ superconductors~\cite{SupraExp}.
These experiments have motivated several theoretical studies of such
physics---these include calculations of such impurity effects in
antiferromagnets~\cite{SpinTh},
superconductors~\cite{SupraTh}, as well as at a
quantum phase transition (QPT) signalling the destruction of antiferromagnetism.\cite{Hoglund_etal_PRL2007}

In this Letter, we use impurities to theoretically probe a class of unconventional QPTs between an
antiferromagnetic phase with long range N\'eel order and a phase with Valence Bond Solid (VBS) order
in square lattice $S=1/2$ magnets. Using an extension of the Sandvik-Evertz valence bond
projector loop Quantum Monte Carlo (QMC)
technique~\cite{Sandvik_Evertz}  that two of us have developed recently~\cite{Banerjee_Damle_unpublished},
we access the total
spin $1/2$ doublet ground state of the system with a missing-spin defect and compute
the spin texture induced by this impurity at these N\'eel-VBS transitions.
We find that this spin texture {\em does not} obey the universal scaling form expected to hold at
scale-invariant quantum critical points. Furthermore, by identifying the actual logarithmic
form of the impurity scaling violations, we argue that the data does {\em not}
support a first order transition.

Much of our interest in the N\'eel-VBS transition of square lattice antiferromagnets
stems from the seminal work of Senthil {\em et al.}~\cite{Senthil_etal_Science2004} who argued that
this QPT is generically continuous, and admits a natural description in terms of `deconfined' strongly interacting $S=1/2$ spinon excitations rather
than the order parameter fields of conventional Landau theory---in this sense, it falls outside the well-known
Landau classification of phase transitions, which predicts a first-order QPT.
Numerical evidence for this theoretical proposal of `deconfined criticality' is
mixed: while Sandvik,\cite{Sandvik_PRL2007,Lou_Sandvik_Kawashima_PRB2009}
and Melko and Kaul~\cite{Melko_Kaul_PRL2008} see an apparently continuous
transition  consistent with deconfined
criticality~\cite{Motrunich_Vishwanath_arXiv2008} in a microscopic S=1/2 spin model, Jiang {\em et al.}~\cite{Jiang_etal_JStatMech2008} provide a detailed analysis
consistent with the competing scenario of a conventional weakly-first order
transition\cite{Kuklov_etal_PRL2008}.

Our identification of logarithmic violations of impurity
scaling at these N\'eel-VBS transitions should be contrasted with the near-perfect scaling collapse
we observe for the spin texture at a different {\em conventional} $T=0$ critical point between a N\'eel ordered
antiferromagnet and a quantum paramagnet {\em without} spontaneous VBS order. These results are
expected to yield important clues about the correct theory of these unconventional N\'eel-VBS transitions.

\begin{figure*}[!]
{\includegraphics[width=2.1\columnwidth]{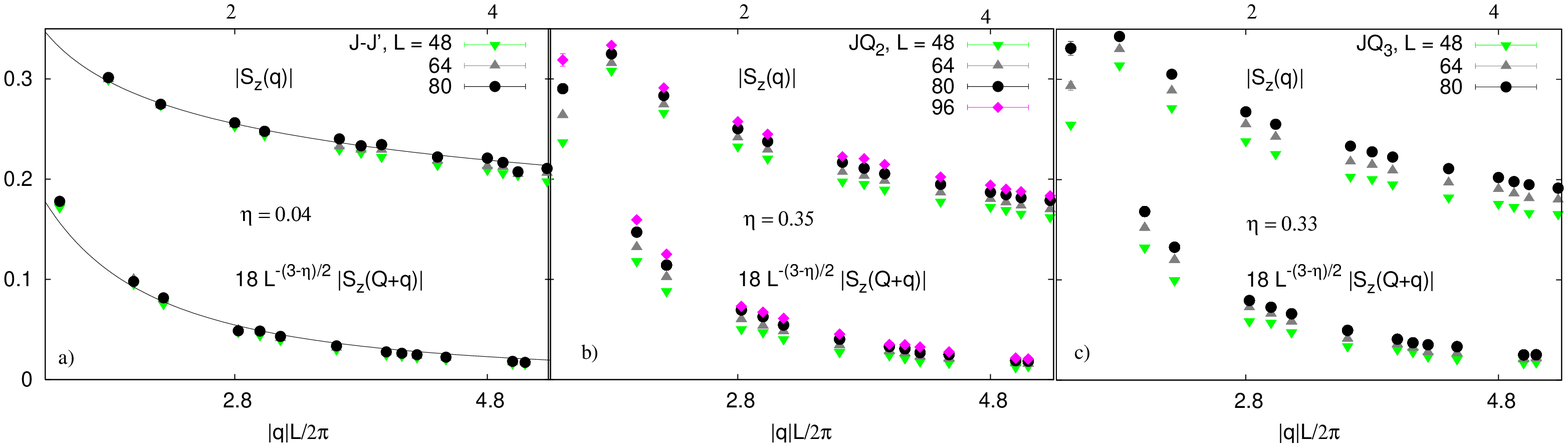}}
\caption{${\mathbf k}$ dependence of $|S_z({\mathbf k})|$ near the ${\mathbf k} = 0$ and ${\mathbf k} = {\mathbf Q}$
peaks at the (a) N\'eel-paramagnet and (b)-(c) N\'eel-VBS transitions. Values of the bulk exponent $\eta$
were taken from Ref.~{\protect{\onlinecite{Hoglund_etal_PRL2007}}} for the N\'eel-paramagnet
transition and from Ref.~{\protect{\onlinecite{Lou_Sandvik_Kawashima_PRB2009}}} for the N\'eel-VBS transitions.
Solid lines in a) are fits to power-law forms obtained by using the value
$\eta^{'} \approx 0.44 \pm 0.02$ for the impurity exponent,
consistent with the estimate in
Ref~{\protect{\onlinecite{Hoglund_etal_PRL2007}}}.
} 
\label{Fig1}
\end{figure*}

We focus here on two putative realizations~\cite{Sandvik_PRL2007,Lou_Sandvik_Kawashima_PRB2009}
of deconfined criticality corresponding to the Hamiltonians ${\mathcal H}_{JQ2}$ and ${\mathcal H}_{JQ3}$ defined
by Sandvik and coworkers:
\be
{\mathcal H}_{JQ2} = -J\sum_{\langle i j \rangle} P_{ij} -Q\sum_{\langle i j \rangle \langle k l \rangle} P_{ij}P_{kl} \nonumber \\
{\mathcal H}_{JQ3} = -J\sum_{\langle i j \rangle} P_{ij} -Q\sum_{\langle i j \rangle \langle k l \rangle \langle r s \rangle} P_{ij}P_{kl}P_{rs} \nonumber
\ee
Here, $P_{ij} = 1/4 - {\mathbf S}_i \cdot {\mathbf S}_j$, $\langle i j \rangle$
refers to a nearest neighbour (n.n.) bond on the square lattice connecting sites $i$ and $j$, and $\langle i j \rangle\langle k l \rangle$
($\langle i j \rangle \langle k l \rangle \langle r s \rangle$) refer
to two (three) adjacent parallel n.n. bonds.
As a foil of the unconventional physics of these JQ models, we also
study a coupled spin-dimer Hamiltonian ${\mathcal H}_{JJ^{'}}$ with
antiferromagnetic n.n. Heisenberg exchange couplings $J$ for all vertical
bonds, and $J$ ($J^{'}$) for even (odd) columns of horizontal bonds~\cite{Wenzel_Janke_PRB2009}.
These models capture two different mechanisms for destabilizing the N\'eel ordered antiferromagnet:
while large values of $Q$ favour a VBS phase in the $JQ_2$ and $JQ_3$ models, large values
of $J^{'}$ drive the system to a quantum paramagnetic state that has no spontaneous symmetry breaking.

In order to study the impurity physics at these transitions, one needs to access the
total spin $1/2$ doublet ground state of an $L \times L$ periodic system with one missing site (periodic boundary
conditions fix $L$ to be even). We have adapted~\cite{Banerjee_Damle_unpublished} the valence bond projector loop-QMC method~\cite{Sandvik_Evertz} to enable an efficient computation of the properties
of ground states with $S_{{\mathrm{tot}}}=1/2$, and $S^{z}_{{\mathrm{tot}}}$ fixed from the
outset to be either $+1/2$ or $-1/2$. This extension of the projector algorithm performs
as well in the $S_{{\mathrm{tot}}} = 1/2$ sector as the original algorithm
does in the singlet sector. Using this modified algorithm, we study the $S_{{\mathrm{tot}}} = S^z_{{\mathrm{tot}}}=1/2$
ground states $|G\rangle$ of periodic systems with a missing spin at ${\mathbf r}= 0$ at the N\'eel-VBS transitions in ${\mathcal H}_{JQ2}$ (at $q_c \equiv (Q/J)_c/((Q/J)_c + 1) \approx 0.962$) and ${\mathcal H}_{JQ3}$ (at $q_c \approx 0.603$)~\cite{Lou_Sandvik_Kawashima_PRB2009},
and at the N\'eel-paramagnet transition of ${\mathcal H}_{JJ^{'}}$ (at $(J^{'}/J)_c \approx 1.9096$)~\cite{Wenzel_Janke_PRB2009}.

The total $S^z = 1/2$ carried by the ground state spreads out throughout the sample to form
the impurity-induced spin texture $\Phi({\mathbf r}) = \langle G| S^z({\mathbf r})|G \rangle$. This texture is expected to have
a smooth uniform part $\Phi^u({\mathbf r})$, and a N\'eel component
$\Phi^n({\mathbf r})$ that alternates in sign between the two sublattices
of the square lattice.

If the QPT in question obeys standard scaling theory, one expects~\cite{Hoglund_etal_PRL2007,Metlitski_Sachdev_PRB2007}
$\Phi^u({\mathbf r}) = \frac{1}{L^2}f^u({\mathbf r}/L)$ and $\Phi^n({\mathbf r}) = \frac{1}{L^{(1+\eta)/2}}f^n({\mathbf r}/L)$, where $\eta$ is the bulk anomalous exponent associated with the N\'eel order parameter, and $f^u$
and $f^n$ are the scaling forms for the uniform and alternating signals.

Earlier work~\cite{Hoglund_etal_PRL2007} has validated this scaling ansatz for
the conventional N\'eel-paramagnet QPT by studying two coarse-grained fields (representing the uniform and alternating signals) obtained from the computed texture $\Phi({\mathbf r})$ by a specific choice of coarse-graining procedure. Although straightforward to implement, such a procedure is somewhat ad-hoc,
and depends on the choice of coarse-graining prescription.

Here we finesse this difficulty by noting that
the Fourier transform $S_z({\mathbf k}) = \sum_{{\mathbf r}}\Phi({\mathbf
  r})\exp(i {\mathbf k} \cdot {\mathbf r})$ (with ${\mathbf k} = 2 \pi
{\mathbf m}/L$ and ${\mathbf m} \equiv (m_x, m_y)$ with integers $m_{x/y} =
0,1, \dots L-1$) is expected to have two peaks, one at ${\mathbf k} = 0$ with magnitude constrained
to be $1/2$, and the second one at ${\mathbf k} = {\mathbf Q} \equiv (\pi,\pi)$ reflecting
the tendency to N\'eel order. The standard scaling ansatz~\cite{Hoglund_etal_PRL2007,Metlitski_Sachdev_PRB2007}
implies that these peaks should satisfy the scaling laws
\be
S_z({\mathbf q}) & = & g_0(L{\mathbf q}) \; \; {\mathrm{for}} \; |{\mathbf q}| \ll \pi/2 \nonumber \\
S_z({\mathbf Q} + {\mathbf q}) &= &L^{(3-\eta)/2}g_{{\mathbf Q}}(L{\mathbf q}) \; \; {\mathrm{for}} \; |{\mathbf q} \ll \pi/2 
\label{qspaceansatz}
\ee
The advantage of this new ${\mathbf k}$ space formulation is clear: one may {\em unambiguously test for scaling
by simply examining} the computed $S_z({\mathbf k})$ for ${\mathbf k}$ in the vicinity of ${\mathbf k} = {\mathbf Q}$ and ${\mathbf k} = 0$. In particular, the data for $S_z({\mathbf q})$ at the transition point computed from samples of varying
size $L$ must fall on top of each other for $|{\mathbf q}| \ll \pi/2$. This
is a completely unbiased test of scaling as it does not need any a
priori estimate of the bulk anomalous exponent $\eta$ for the N\'eel
order parameter, nor does it rely on a specific coarse-graining procedure to
define the scaling components of the texture.

\begin{figure*}[!]
{\includegraphics[width=2\columnwidth]{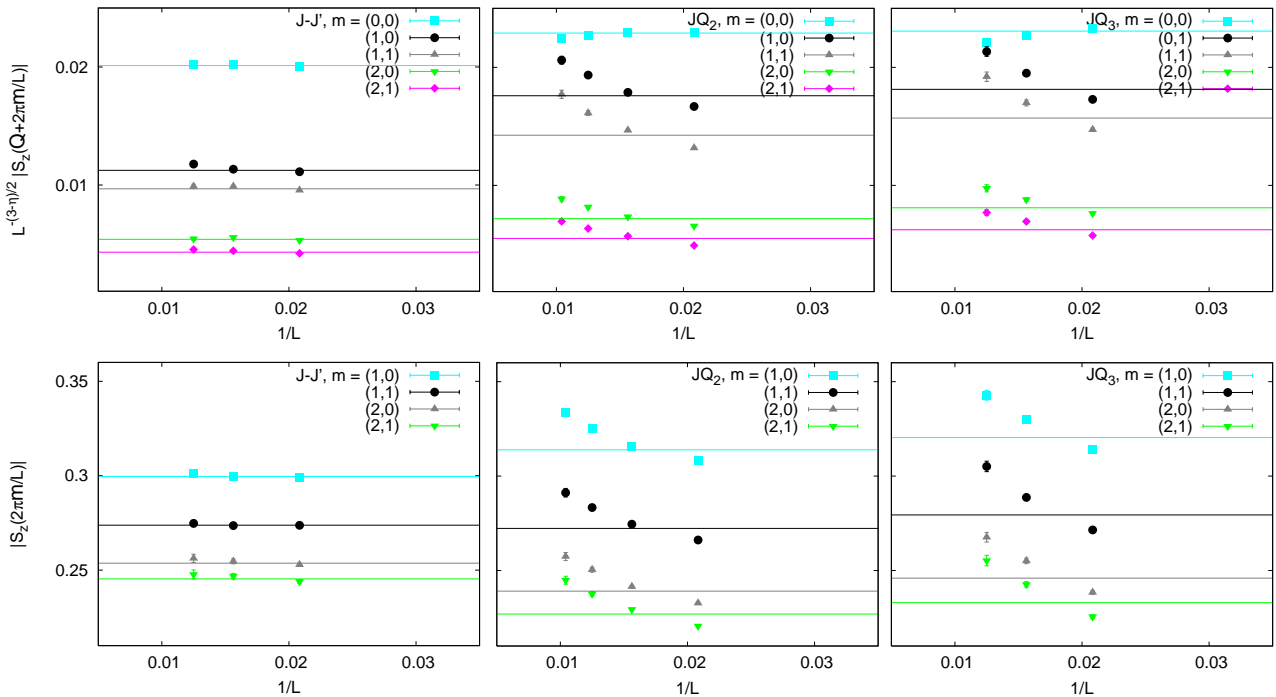}}
\caption{$1/L$ dependence of $|S_z({\mathbf Q}+2 \pi{\mathbf m}/L)|/L^{(3-\eta)/2}$ and $|S_z(2 \pi{\mathbf m}/L)|$ for small $|{\mathbf m}|$ at the N\'eel-VBS
and N\'eel-Paramagnet transitions. All horizontal lines are guides to the eye that indicate the expected
behaviour if scaling was perfectly obeyed, and values of $\eta$ are the same as in Fig~{\protect{\ref{Fig1}}}.}
\label{Fig2}
\end{figure*}

Our first inkling that standard scaling does not work at
these N\'eel-VBS transitions comes from the computed
values of $|S_z({\mathbf q})|$ shown in Fig~\ref{Fig1} and Fig~\ref{Fig2} for ${\mathbf q} = 2
\pi {\mathbf m}/L$ with $|{\mathbf m}| \ll L/2$ (in practice, we focus on
$|{\mathbf m}| \lesssim L_{{\mathrm{min}}}/12$ and average over all ${\mathbf m}$ that correspond to a given $|{\mathbf m}|$). Larger values of $L$
are seen to yield a systematically larger value of $|S_z|$ at the same $|{\mathbf m}|$. This behaviour at
the N\'eel-VBS transitions is in clear violation of the scaling form
Eq.~\ref{qspaceansatz}; this should be contrasted with the excellent scaling observed at the conventional N\'eel-paramagnet critical point of the $J-J^{'}$ model. 
Given the unbiased nature of this test of scaling, we consider this rather strong evidence
for violation of impurity scaling properties at these N\'eel-VBS transitions in the $JQ_2$ and $JQ_3$
models.

Next, we analyze the Bragg peak at the antiferromagnetic
wavevector, ${\mathbf k} = {\mathbf Q}$, focusing on the $L$ dependence
at the N\'eel-VBS transition point in both $JQ$ models. We find no evidence of any double-peak
structure for the histogram of $|S_z({\mathbf Q})|$ at these N\'eel-VBS transitions, implying that the first order
jump in the order parameter, if any, is immeasurably small even at sizes as large as $L=96$.
Furthermore, we confirm that the computed
values obey the power-law scaling $|S_z({\mathbf Q})| \sim L^{(3-\eta)/2}$ quite well at both N\'eel-VBS transitions,
with the anomalous exponents $\eta_{JQ_3} \approx 0.33$ and  $\eta_{JQ_2} \approx 0.35$ taken
from Ref~\onlinecite{Lou_Sandvik_Kawashima_PRB2009,Melko_Kaul_PRL2008}.
Our results for the $J-J^{'}$ model are also consistent with the power-law scaling $|S_z({\mathbf Q})| \sim L^{(3-\eta)/2}$ with
the known value of $\eta \approx 0.04$ for the N\'eel-paramagnet QPT~\cite{Hoglund_etal_PRL2007}.

However, violations of impurity scaling in the staggered component of the
texture at the N\'eel-VBS transitions become evident when one tests for scaling collapse
at ${\mathbf k} = {\mathbf Q} + 2\pi {\mathbf m}/L$ with $|{\mathbf m}|$ small but non-zero (we focus on
$|{\mathbf m}| \lesssim L_{{\mathrm{min}}}/12$ and average over all ${\mathbf m}$ that correspond to a given $|{\mathbf m}|$)
---again, larger $L$ give larger values of $|S_z({\mathbf k})|$ for
the same non-zero $|{\mathbf m}|$(Fig~\ref{Fig1} and Fig~\ref{Fig2}). This is again underlined by the excellent scaling collapse
exhibited by the corresponding quantities computed at the N\'eel-paramagnet quantum critical point
of the $J-J^{'}$ model (Figures~\ref{Fig1} and~\ref{Fig2}).

\begin{figure}[h]
{\includegraphics[width=\columnwidth]{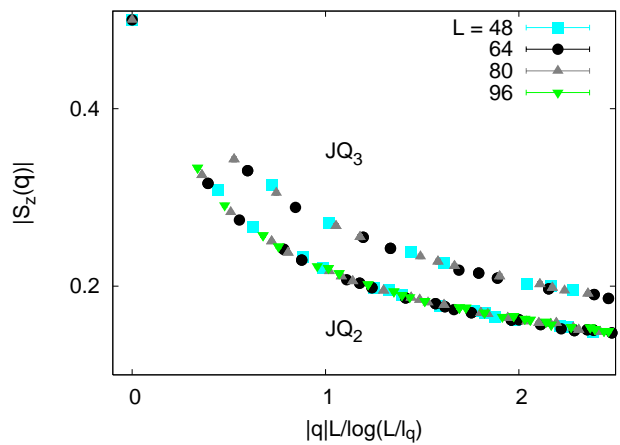}}
\caption{Logarithmically modified scaling collapse of $S_z({\mathbf k})$ near ${\mathbf k} = 0$
at the N\'eel-VBS transition, with $l_0 = 5\pm1 $ ($l_0 = 12\pm1$) for the $JQ_2$ ($JQ_3$) model.}
\label{Fig3}
\end{figure}
\begin{figure}[h]
{\includegraphics[width=\columnwidth]{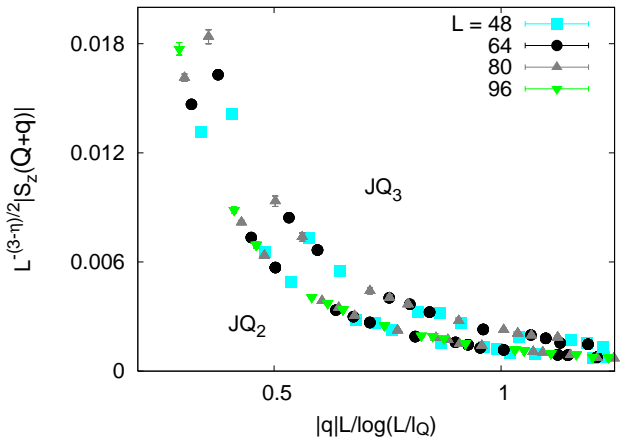}}
\caption{Logarithmically modified scaling collapse of $S_z({\mathbf k})$ near ${\mathbf k} = {\mathbf Q}$
at the N\'eel-VBS transition,  with $l_{\mathbf Q} = 0.75\pm0.2  $ ($l_{\mathbf Q} = 1.5\pm0.5$) for the $JQ_2$ ($JQ_3$) model.}
\label{Fig4}
\end{figure}

What do these violations of quantum-critical scaling signal? This is best addressed by asking
if the computed spin texture satisfies some suitably modified scaling laws. To explore
this, we first note that the absence of any power-law prefactor to $g_{0}$ in
the scaling ansatz Eq.~\ref{qspaceansatz} reflects
the conservation of total spin and the specific power of $L$ that multiplies $g_{{\mathbf Q}}$
reflects the presence of power-law N\'eel order at criticality, while the scaling
argument $L{\mathbf q}$ of both functions follows simply from the statement
that $L$ is the only length scale of relevance to the long-distance physics
at a scale invariant critical point.

This form of the scaling argument can break down if the effective low-energy theory has a term which is {\em marginally}
irrelevant for the long-distance physics, and renormalizes to zero slowly. This can introduce
an additional length scale, and give rise to
logarithmic violations of scaling.
Indeed, signatures of such logarithmic drifts have been recently seen by Sandvik
in his analysis of 
the bulk N\'eel-VBS transitions~\cite{Sandvik_latest}.

This motivates us to ask if the computed spin texture obeys
a modified scaling form (for $|{\mathbf q}| \ll \pi/2$)
\be
 S_z({\mathbf q}) & = & g_0(L{\mathbf q}/\log(L/l_0)) \nonumber \\
S_z({\mathbf Q} + {\mathbf q}) &= &L^{(3-\eta)/2}g_{{\mathbf Q}}(L{\mathbf q}/\log(L/l_{{\mathbf Q}}))  
\label{logansatz}
\ee
where $l_0$ and $l_{{\mathbf Q}}$ are related to the additional non-universal length scale
introduced by the slow vanishing of a marginally irrelevant term in the effective Hamiltonian.
As is clear from Figs~\ref{Fig3}~\ref{Fig4}, the answer is yes: this modified scaling law
gives an extremely good account of our results.

These logarithmic violations of scaling are at odds with predictions of the theory
of deconfined criticality~\cite{Metlitski_Sachdev_PRB2007}. However, given
the absence of any clear signal for phase-coexistence at the transition
point, one cannot simply ascribe these violations to the presence of an immeasurably
weak first-order jump in the order parameters. We are thus led to conclude that although the N\'eel-VBS
transition is continuous, the theory of deconfined criticality~\cite{Senthil_etal_Science2004} needs to be modified in order to
account for these logarithmic corrections. This conclusion underscores the utility of impurity physics
as a probe of complex strongly-correlated states of many-body systems. An
interesting follow-up would be to use the same probe at non-zero temperature above the QPT and
test for violations of scaling predictions for the impurity susceptibility.

We thank M.~Metliski and S.~Sachdev for clarifications regarding their results~\cite{Metlitski_Sachdev_PRB2007}, A.~W.~Sandvik for clarifications
regarding Ref~\onlinecite{Hoglund_etal_PRL2007} and communicating his recent results~\cite{Sandvik_latest} prior to submission, S.~Chandrashekaran for
clarifications regarding Ref~\onlinecite{Jiang_etal_JStatMech2008},
S.~Capponi for help in benchmarking our QMC method, and S.N.~Majumdar for a critical reading of this draft. We acknowledge computational
resources of TIFR and funding from DST-SR/S2/RJN-25/2006 (KD).


\begin{thebibliography}{999}

\bibitem{Senthil_etal_Science2004} T. Senthil {\it et al.}, Science {\bf 303}, 1490 (2004).

\bibitem{Alloul_etal_RMP2009}  H.~Alloul, J.~Bobroff, M.~Gabay, and P.~J.~Hirschfeld, Rev. Mod. Phys. {\bf 81}, 45 (2009).


\bibitem{SpinExp} M. Takigawa {\it et al.},  Phys. Rev. B {\bf 55}, 14129
  (1997); J.~Bobroff {\it et al.}, Phys. Rev. Lett. {\bf 103}, 047201
  (2009); F.~Tedoldi, R.~Santachiara, and M.~Horvatic, {\it ibid} {\bf 83},
  412 (1999); J.~Das {\it et al.}, Phys. Rev. B {\bf 69}, 144404 (2004).

\bibitem{Sachdev_etal_Science1999} S.~Sachdev, C.~Buragohain, and M.~Vojta, Science {\bf 286}, 2479 (1999).

\bibitem{SupraExp} A. V. Mahajan, H. Alloul, G. Collin, and
  J. F. Marucco, Phys. Rev. Lett. {\bf 72}, 3100 (1994); S.H.~Pan {\it et
    al.}, Nature {\bf 403}, 746 (2000); J.~Bobroff {\it et al.},  Phys. Rev. Lett. {\bf 83},
  4381 (1999); {\it ibid}  {\bf 86}, 4116 (2000).

\bibitem{SpinTh} G.B. Martins {\it et al.}, Phys. Rev. Lett. {\bf 78}, 3563
  (1997); S. Eggert {\it et al.}, {\it ibid} {\bf 99}, 097204 (2007);
  R.~K.~Kaul, R.~G.~Melko, M.~A.~Metlitski, and S.~Sachdev, {\it ibid} {\bf 101}, 187206 (2008).

\bibitem{SupraTh} A.~Polkovnikov, S.~Sachdev, and M.~Vojta,
  Phys. Rev. Lett. {\bf 86}, 296 (2001); M. Capello and D. Poilblanc, Phys. Rev. B {\bf 79}, 224507 (2009).

\bibitem{Hoglund_etal_PRL2007} K.~H.~Hoglund, A.~W.~Sandvik, and S.~Sachdev, Phys. Rev. Lett. {\bf 98}, 087203 (2007).

\bibitem{Sandvik_Evertz} A.W.~Sandvik and H.-G.~Evertz, arXiv:0807.0682, unpublished; A.W.~Sandvik, Phys. Rev. Lett. {\bf 95}, 207203 (2005).

\bibitem{Banerjee_Damle_unpublished} A.~Banerjee and K.~Damle, unpublished.

\bibitem{Sandvik_PRL2007} A.~W.~Sandvik, Phys. Rev. Lett. {\bf 98}, 227202 (2007).

\bibitem{Lou_Sandvik_Kawashima_PRB2009} J.~Lou, A.~W.~Sandvik, and
  N.~Kawashima, Phys. Rev. B {\bf 80}, 180414 (2009)

\bibitem{Melko_Kaul_PRL2008} R.~G.~Melko and R.~K.~Kaul, Phys. Rev. Lett. 100, 017203 (2008).

\bibitem{Motrunich_Vishwanath_arXiv2008} O.~I.~Motrunich and A.~Vishwanath, arXiv:0805.1494, unpublished.

\bibitem{Jiang_etal_JStatMech2008} F.~J.~Jiang, M.~Nyfeler, S~Chandrasekharan, and U.~J.~Wiese, J. Stat. Mech. P02009 (2008).

\bibitem{Kuklov_etal_PRL2008} A. B. Kuklov {\it et al.},  Phys. Rev. Lett. {\bf 101}, 050405 (2008).

\bibitem{Wenzel_Janke_PRB2009} S.~Wenzel and W.~Janke, Phys. Rev. B {\bf 79}, 014410 (2009). 

\bibitem{Metlitski_Sachdev_PRB2007} M.~Metlitski and S.~Sachdev, Phys. Rev. B {\bf 76}, 064423 (2007).

\bibitem{Sandvik_latest} A.~W.~Sandvik, arXiv:1001.4296, (unpublished).
\end{thebibliography}
\end{document}